\documentclass[aps,prl,showpacs,twocolumn]{revtex4}
\usepackage{amssymb}
\usepackage{epsfig,graphicx}

\begin{document}

\title{Internal Josephson-Like Tunneling in Two-Component Bose-Einstein
Condensates Affected by Sign of the Atomic Interaction and External
Trapping Potential}
\author{Bo Xiong, X. X. Liu}
\affiliation{Beijing National Laboratory for Condensed Matter Physics, Institute of
Physics, Chinese Academy of Sciences, Beijing 100080, China}
\date{\today}

\begin{abstract}
We study the Josephson-like tunneling in two-component Bose-Einstein
condensates coupled with microwave field in respond to various
attractive and repulsive atomic interaction under the various aspect
ratio of trapping potential and the gravitational field. It is very
interesting to find that the dynamic of Josephson-like tunneling can
be controlled from fast damped oscillations and asymmetric
occupation to nondamped oscillation and symmetric occupation.
\end{abstract}

\maketitle

The existence of a Josephson current is a direct manifestation of
macroscopic quantum phase coherence \cite{Barone1982} and have numerous
important applications in condensed matter physics, quantum optics and cold
atom physics, for example, precision measurement, quantum computation. The
physical origin of the Josephson current is the temporal interference of the
two systems which must both have a well defined quantum phase and a
different average energy per particle, respectively. Recently, the
experimental realization of multi-component Bose-Einstein condensates (BECs)
of weakly interacting alkali atoms has provided a route to study\ Josephson
effect \cite%
{Hall1998-1,MatthewsPRL1998,LeggetRMP,AndersonScience,WilliamsPRA,ParkPRL,CataliottiScience,KohlerPRL,ShinPRL}
in a controlled and tunable way by means of Feshbach resonance \cite%
{SaitoPRL,KevrekidisPRL,PelinovskyPRL,Perez-GarciaPRL,KonotopPRL,MatuszewskiPRL,Muryshev,InouyeNature,RobertsPRL,StreckerNature}
so far unattainable in charged systems. The physical origin of the
low-energy Feshbach resonances is the\textbf{\ }low-energy binary collisions
described by the difference of the initial and intermediate state energies
which can be effectively altered through variations of the strength of an
external magnetic field.

In this Letter, we study the dynamics of Josephson-like tunneling in
two-component Bose-Einstein condensates coupled with microwave field under
tuning the atomic interaction from attractive case to repulsive case in the
various aspect ratio of trapping potential. Through changing the relative
value between attractive intra-atomic and inter-atomic interaction, we find
that, occupation on relative number of atoms represents two very different
behavior, which one is fast damped oscillation, but the other is nondamped
and irregular oscillation. In particular, when the atomic interaction is
changed from attractive to repulsive case, we find that relative number of
atoms also represents nondamped and irregular oscillation.
\begin{figure}[tbp]
\epsfxsize=9.5cm \centerline{\epsffile{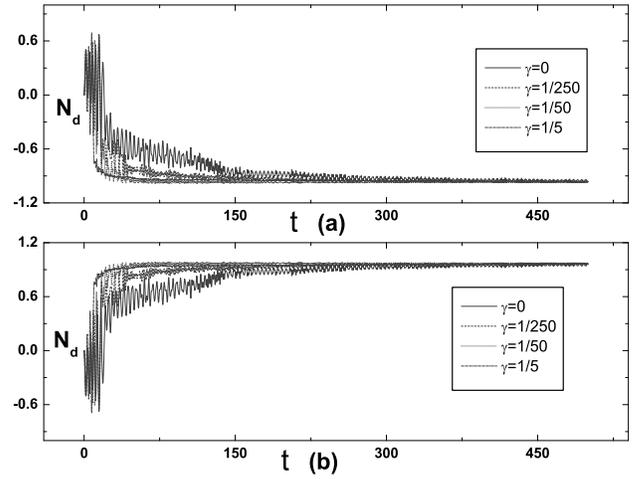}}
\caption{Temporal evolution of relative number of atoms $N_{d}$ for initial
Gaussian wave function in response to various aspect ratio of trapping
potential $\protect\gamma $. (a) Corresponding to attractive atomic
interactions for the parameters as $G_{11}=G_{22}=4G_{12}=-10$ and $\Omega
=1.5\protect\omega _{\bot }$, $\protect\delta =0.5\protect\omega _{\bot }$.
(b) Corresponding to attractive atomic interactions for the parameters as $%
G_{11}=G_{22}=4G_{12}=-10$ and $\Omega =1.5\protect\omega _{\bot }$, $%
\protect\delta =-0.5\protect\omega _{\bot }$. }
\end{figure}

We consider a system of longitudinal elongated two-component BECs coupled by
the microwave field with the effective Rabi frequency $\Omega $ and finite
detune $\delta $, where the $\Omega $ and $\delta $ are independent upon
time and space coordinate as experimental case \cite{MatthewsPRL1998}. At
zero temperature, this system can be described by macroscopic wave function $%
\psi _{j}$\ (normalized to unity, i.e., $\int (|\psi _{1}|^{2}+|\psi
_{2}|^{2})dz=1$) which obey the dimensionless one-dimensional
Gross-Pitaevskii equations (GPEs) as \cite{WilliamsPRA,ParkPRL}%
\begin{eqnarray}
i\frac{\partial }{\partial t}\psi _{j}(z,t) &=&[-\frac{\partial ^{2}}{%
\partial z^{2}}+V(z)]\psi _{j}(z,t)+  \nonumber \\
&&[G_{jj}(t)|\psi _{j}|^{2}+G_{jk}(t)|\psi _{k}|^{2}]\psi _{j}(z,t)+
\nonumber \\
&&(-1)^{j+1}\frac{\delta }{\omega _{\bot }}\psi _{j}(z,t)+\frac{\Omega }{%
\omega _{\bot }}\psi _{k}(z,t),  \label{GP2}
\end{eqnarray}%
where $j,k=1,2$, $j\neq k$, and $\omega _{\bot }$ is radial harmonic
trapping frequency, time and coordinate unit are $2/\omega _{\bot }$ and $%
a_{\bot }=\sqrt{\hbar /m\omega _{\bot }}$, respectively, and $V(z)$ is the
longitudinal confining potential, $G_{jj}=(4N\hbar a_{jj})/(ma_{\bot
}^{3}\omega _{\bot })$ and $G_{jk}=(4N\hbar a_{jk})/(ma_{\bot }^{3}\omega
_{\bot })$ are dimensionless intra- and inter-species effective atomic
interaction potentials, respectively, where $a_{jj}$ and $a_{jk}$ are
corresponding s-wave scattering lengths, respectively, $N$ is the total
atomic number, and $m$ is the atomic mass assumed the same for both two
species.
\begin{figure}[tbp]
\epsfxsize=9.5cm \centerline{\epsffile{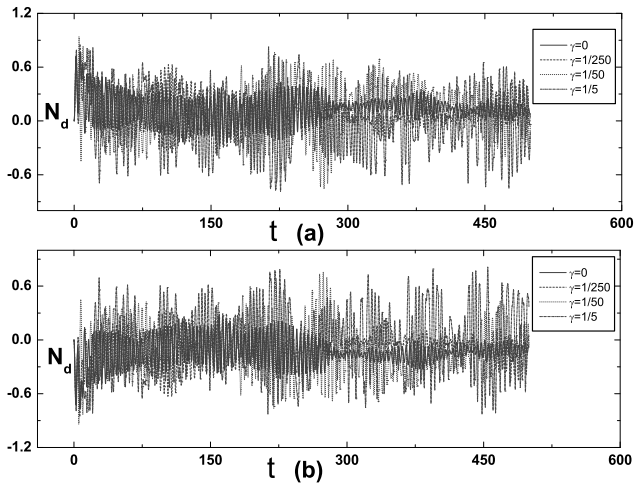}}
\caption{Temporal evolution of relative number of atoms $N_{d}$ for initial
Gaussian wave function in response to various aspect ratio of trapping
potential $\protect\gamma $. (a) Corresponding to attractive atomic
interactions for the parameters as $4G_{11}=4G_{22}=G_{12}=-10$ and $\Omega
=1.5\protect\omega _{\bot }$, $\protect\delta =0.5\protect\omega _{\bot }$.
(b) Corresponding to attractive atomic interactions for the parameters as $%
4G_{11}=4G_{22}=G_{12}=-10$ and $\Omega =1.5\protect\omega _{\bot }$, $%
\protect\delta =-0.5\protect\omega _{\bot }$. }
\end{figure}

In the following discussion, we consider two-hyperfine spin states of $%
^{7}Li $ bose atom which the magnitude and sign of the atomic interaction
can be tuned to any value via a magnetic-field dependent Feshbach resonance.
And we choose dimensionless $V(z)=\gamma ^{2}z^{2}+\lambda z$, where $\gamma
=\omega _{z}/\omega _{\bot }$ is aspect ratio of trapping potential and $%
\lambda $ determines the strength of the gravitational field, $\omega _{z}$
is longitudinal harmonic trapping frequency and $\omega _{\bot }=2\pi \times
625$ Hz (so, the time and coordinate units are $0.51$ ms and $a_{\bot }=1.51$
$\mu $m respectively). Moreover, considering the condensates will be
unstable and collapse when the number of particles is large enough \cite%
{Muryshev}, so we choose $N=6000$, which provides a safe range of parameters
for avoiding instability and collapse occurring.

First of all, we consider the attractive case, i.e., intra- and
inter-species effective atomic interaction potentials are\ time-independent
attractive case, which can be easily realized using Feshbach resonance
experimentally. And the effective Rabi frequency and finite detune are
chosen as $\Omega =1.5\omega _{\bot }$ and $\delta =0.5\omega _{\bot }$
corresponding to Fig. 1a and $\delta =-0.5\omega _{\bot }$ corresponding to
Fig. 1b. Here, we have chosen $a_{11}=a_{22}$ is just considering the
experimental feasibility and assumed that tuning atomic interaction wouldn't
destroy the condensates in this system. We denote relative number of atoms
between two species as $N_{d}$, i.e., $N_{d}=N_{1}-N_{2}$, where $N_{1}=\int
|\psi _{1}|^{2}dz$ and $N_{2}=\int |\psi _{2}|^{2}dz$, respectively. We
study Josephson-like tunneling in this case and the results are shown in
Fig. 1, which is based on numerical simulation of Eqs. (\ref{GP2}) for the
initial Gaussian wave function under the various aspect ratio of trapping
potential. As shown in Fig. 1, during the evolution process, exchange of
atoms is always damped very fast and almost suppressed completely
irrespective of aspect ratio of trapping potential. In particular, it is
important to notice that occupation on the each components represents very
asymmetric behavior and the atoms almost transform to second component
completely in the case of $\delta =0.5\omega _{\bot }$, but almost transform
to first component completely in the case of $\delta =-0.5\omega _{\bot }$.
In this case, system has lower energy through exchanging the atoms from
first to second components under the condition $\delta =0.5\omega _{\bot }$,
but from second to first components under the condition $\delta =-0.5\omega
_{\bot }$.
\begin{figure}[tbp]
\epsfxsize=9.5cm \centerline{\epsffile{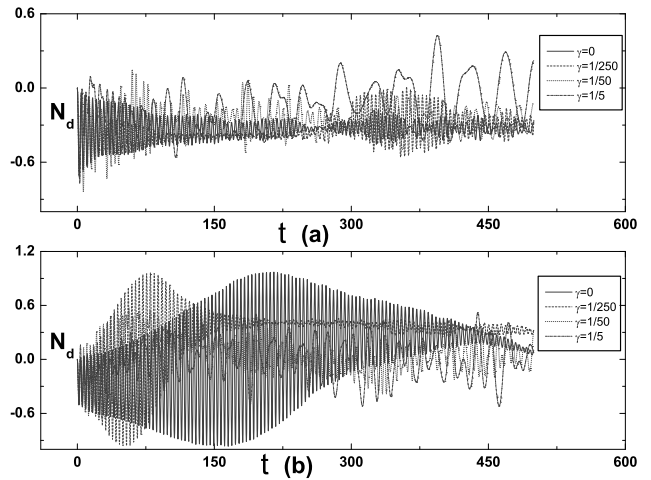}}
\caption{Temporal evolution of relative number of atoms $N_{d}$ for initial
Gaussian wave function in response to various aspect ratio of trapping
potential $\protect\gamma $. (a) Corresponding to repulsive atomic
interactions for the parameters as $G_{11}=G_{22}=4G_{12}=10$ and $\Omega
=1.5\protect\omega _{\bot }$, $\protect\delta =-0.5\protect\omega _{\bot }$.
(b) Corresponding to repulsive atomic interactions for the parameters as $%
4G_{11}=4G_{22}=G_{12}=10$ and $\Omega =1.5\protect\omega _{\bot }$, $%
\protect\delta =-0.5\protect\omega _{\bot }$. }
\end{figure}

From the experimental points of view, it is very important to study the
dynamics of Josephson-like tunneling under the tuning atomic interaction
which provides a route to study\ Josephson effect in a controlled and
tunable way by means of Feshbach resonance so far unattainable in charged
systems. So, we consider the attractive intra- and inter-species effective
atomic interaction potentials become $4G_{11}=4G_{22}=G_{12}=-10$. The
results are shown in Fig. 2, which is based on numerical simulation of Eqs. (%
\ref{GP2}) for the initial Gaussian wave function under the various aspect
ratio of trapping potential. We can see that amplitude of oscillation
becomes larger with increasing the aspect ratio of trapping potential $%
\gamma $. Especially, contrary to first case, occupation on the each
components almost represents very symmetric behavior and the atoms are not
favorite to stay on only one of two components because the energy of system
in asymmetric occupation is larger than symmetric occupation on the each
components.

Next, we study the dynamics of Josephson-like tunneling under the repulsive
atomic interaction in response to various aspect ratio of trapping
potential. As shown in Fig. 3(a) corresponding to the case $%
G_{11}=G_{22}=4G_{12}=10$ and $\Omega =1.5\omega _{\bot }$, $\delta
=-0.5\omega _{\bot }$, amplitude of oscillation becomes larger with
increasing the aspect ratio of trapping potential $\gamma $, and occupation
on the each components also represents symmetric behavior approximately,
which is like the case shown in Fig. 2. However, oscillation becomes slower
with increasing the aspect ratio of trapping potential $\gamma $, which is
very different from the case shown in Fig. 2. It is very interesting to
notice that Fig. 3(b) corresponding to the case $4G_{11}=4G_{22}=G_{12}=10$
represents very different phenomena from Fig. 3(a).
\begin{figure}[tbp]
\epsfxsize=9.5cm \centerline{\epsffile{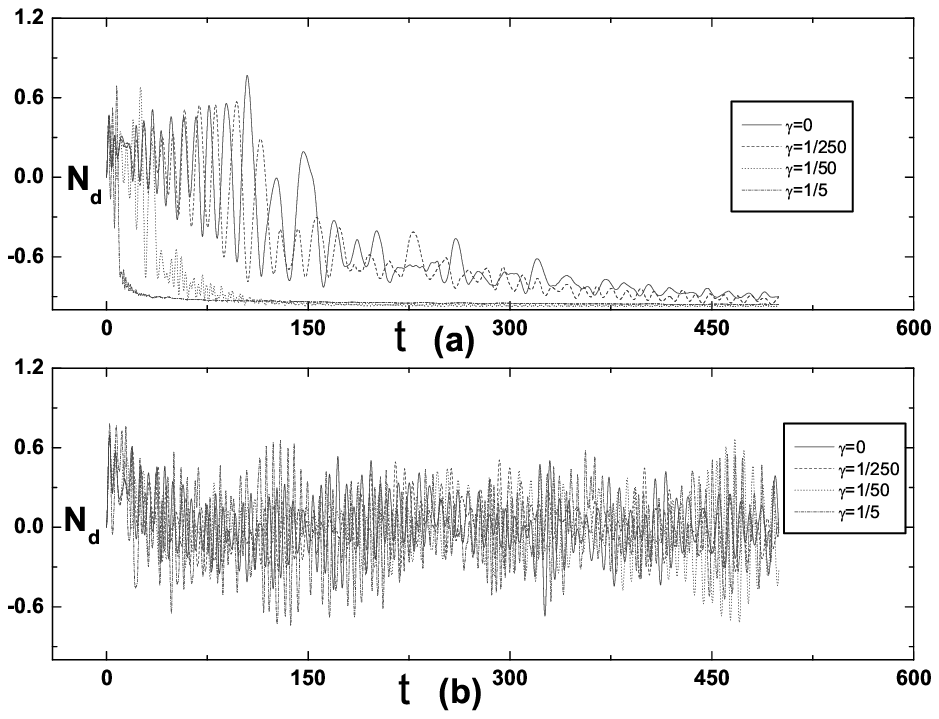}}
\caption{Under the gravitational field, temporal evolution of relative
number of atoms $N_{d}$ for initial Gaussian wave function in response to
various aspect ratio of trapping potential $\protect\gamma $. (a)
Corresponding to attractive atomic interactions for the parameters as $%
G_{11}=G_{22}=4G_{12}=-10$ and $\Omega =1.5\protect\omega _{\bot }$, $%
\protect\delta =0.5\protect\omega _{\bot }$. (b) Corresponding to attractive
atomic interactions for the parameters as $4G_{11}=4G_{22}=G_{12}=-10$ and $%
\Omega =1.5\protect\omega _{\bot }$, $\protect\delta =0.5\protect\omega %
_{\bot }$. }
\end{figure}

Another interesting case is to study the dynamics of Josephson-like
tunneling under the gravitational field, so $V(z)\ $in Eqs. (\ref{GP2})
becomes $\gamma ^{2}z^{2}+\lambda z$, where $\lambda $ determines the
strength of the gravitational field and equals $0.839$ in unit of $%
(1/2)\hbar \omega _{\bot }$. As shown in Fig. 4(a) corresponding to
attractive atomic interaction and $G_{11}=G_{22}=4G_{12}=-10$, $\Omega
=1.5\omega _{\bot }$, $\delta =0.5\omega _{\bot }$, we can see that the
relative number of atoms is damped slower than the case without the
gravitational field when $\gamma $ is very small, but is damped with the
same speed as the case without the gravitational field when $\gamma $ is
large as shown in Fig. 1. In Fig. 4(b) corresponding to attractive atomic
interaction and $4G_{11}=4G_{22}=G_{12}=-10$, $\Omega =1.5\omega _{\bot }$, $%
\delta =0.5\omega _{\bot }$, we can see the same phenomena as the case
without the gravitational field as shown in Fig. 2, except the amplitude and
frequency of oscillation of relative number of atoms have some differences.
\begin{figure}[tbp]
\epsfxsize=9.5cm \centerline{\epsffile{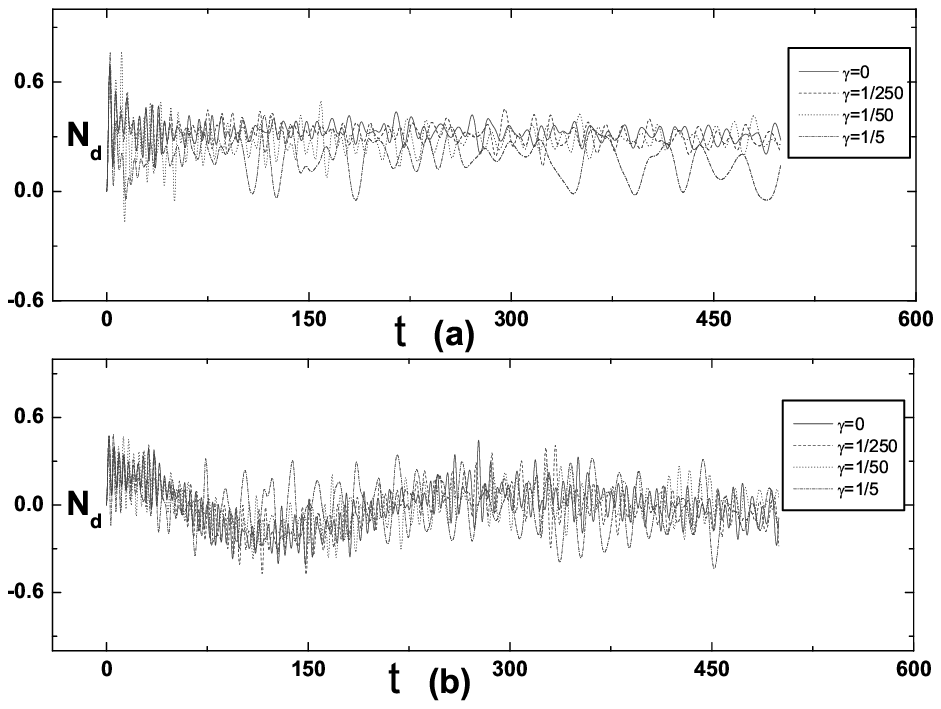}}
\caption{Under the gravitational field, temporal evolution of relative
number of atoms $N_{d}$ for initial Gaussian wave function in response to
various aspect ratio of trapping potential $\protect\gamma $. (a)
Corresponding to attractive atomic interactions for the parameters as $%
G_{11}=G_{22}=4G_{12}=10$ and $\Omega =1.5\protect\omega _{\bot }$, $\protect%
\delta =0.5\protect\omega _{\bot }$. (b) Corresponding to attractive atomic
interactions for the parameters as $4G_{11}=4G_{22}=G_{12}=10$ and $\Omega
=1.5\protect\omega _{\bot }$, $\protect\delta =0.5\protect\omega _{\bot }$. }
\end{figure}

When the atomic interaction becomes repulsive under the gravitational field,
the dynamics of Josephson-like tunneling is shown in Fig. 5. We can see that
under the atomic interaction satisfied $G_{11}=G_{22}=4G_{12}=-10$
corresponding to Fig. 5(a), the dynamics of Josephson-like tunneling under
the gravitational field represents the same phenomena as the case without
the gravitational field as shown in Fig. 3(a), except the amplitude and
frequency of oscillation of relative number of atoms have some differences.
But under the atomic interaction satisfied $4G_{11}=4G_{22}=G_{12}=10$
corresponding to Fig. 5(b) and $\gamma $ is very small, the dynamics of
Josephson-like tunneling under the gravitational field is very different
from the case without the gravitational field as shown in Fig. 3(b).

In conclusion, we find a robust way to control the dynamics of
Josephson-like tunneling from fast damped oscillations and asymmetric
occupation to nondamped oscillation and symmetric occupation in
two-component Bose-Einstein condensates coupled with microwave field. Recent
developments of controlling the scattering length and realization of
multi-component BECs in the experiments allow for the experimental
investigation of our prediction in future.

This work was supported by the NSF of China under 10474055, 90406017,
10588402, 60525417; the NKBRSF of China under 2005CB724508; the STCM of
Shanghai under 05PJ14038, 06JC14026, 04DZ14009.

\end{document}